\documentclass[journal ]{new-aiaa}

\usepackage[utf8]{inputenc}
\usepackage{textcomp}

\usepackage{hyperref}

\usepackage{bm}

\usepackage{easyReviewMod}   

\newcommand{\bmtx}{\begin{bmatrix}}
\newcommand{\emtx}{\end{bmatrix}}
\newcommand{\bsmtx}{\left[ \begin{smallmatrix}} 
\newcommand{\esmtx}{\end{smallmatrix} \right]}
\newcommand{\bmatarray}[1]{\left[\begin{array}{#1}}
\newcommand{\ematarray}{\end{array}\right]}

\newcommand{\abs}[1]{\left|#1\right|}

\newcommand{\E}{\mathbb{E}}

\renewcommand{\d}{\mathrm{d}}

\newcommand{\ve}[1]{\boldsymbol{#1}}

\usepackage{tikz}
\usetikzlibrary{positioning, arrows, calc, shadings}

\tikzstyle{blockdiag}	= [node distance=5mm, >=stealth', semithick]
\tikzstyle{block}			= [draw, rectangle, minimum width=1cm, minimum 
height=.8cm]
\tikzstyle{sum} = [draw,circle,inner sep=0pt, minimum size=6pt]
\tikzstyle{connector} = [draw,circle,inner sep=0pt, minimum size=2pt, 
fill=black]

\usepackage{array}
\newcolumntype{P}[1]{>{\centering\arraybackslash}p{#1}}
\newcolumntype{M}[1]{>{\centering\arraybackslash}m{#1}}
\definecolor{blue1}{RGB}{222,235,247}
\definecolor{blue2}{RGB}{158,202,225}
\definecolor{blue3}{RGB}{49,130,189}

\usepackage{pgfplots}
  \usepgfplotslibrary{groupplots}
  \usepgfplotslibrary{fillbetween}

\usepackage[dvipsnames]{xcolor}

\usepackage{graphicx}
\usepackage{amsmath}
\usepackage[version=4]{mhchem}
\usepackage{siunitx}
\usepackage{longtable,tabularx}
\usepackage{comment}

\def\DS{\displaystyle}

\title{Method to Compute Pointing Displacement, Smear, and Jitter Covariances for Optical Payloads}

\author{Peter Seiler \footnote{Professor, Department of Electrical Engineering and Computer Science, pseiler@umich.edu, AIAA Associate Fellow}}
\affil{University of Michigan, Ann Arbor, Michigan, 48109, United States}

\author{Mark E. Pittelkau\footnote{Independent Consultant, Aerospace Control Systems, LLC, mpittelkau@acsinnovations.com, AIAA Associate Fellow, IEEE Life Senior Member}}
\affil{Aerospace Control Systems, LLC, Round Hill, VA 20141, United States}

\author{Felix Biert\"umpfel \footnote{Marie-Curie Postdoctoral Fellow, Chair of Flight Mechanics and Control and Department of Electrical Engineering and Computer Science, felix.biertuempfel@tu-dresden, felixb@umich.edu, AIAA Young Professional Member (Corresponding Author)}}
\affil{TU Dresden, Dresden, Saxony, 01307, Germany}
\affil{University of Michigan, Ann Arbor, Michigan, 48109, United States}

\begin{document}

\maketitle

\vspace{-10pt}
\section{Introduction}

\lettrine{T}{he} imaging performance of an optical payload is potentially limited by image motion during an exposure. Image motion reduces the modulation transfer function (MTF) --- the spatial frequency response --- of the optical system, thus adversely affecting image quality \cite{pittelkau16}. The image motion is due to bus and optical payload pointing control system errors driven by various noise and disturbance sources. See, for example, \cite{dennehy19} for a survey of noise and disturbance sources on spacecraft. Here, the ``bus'' refers to any kind of platform that carries an optical payload, including spacecraft, aircraft (planes, drones, balloons), ships, ground vehicles, robots, and even hand-held cameras. In this work, the optical system is a isoplanatic (shift-invariant) noncoherent (incoherent) imaging system.

Components of image motion over an exposure interval include displacement (shift), smear (linear), jitter (Gaussian blur), smile (quadratic), frown (cubic), exponential decay, tonal and multi-tonal motion, and fractional cycles. Tonal and multi-tonal motion may be included in jitter under certain conditions \cite{pittelkau2023}. 
In previous work \cite{lucke92, sirlin90, pittelkau03, bayard04}, ``jitter'' includes image smear motion. Smear and jitter are treated separately in \cite{pittelkau16} due to their different effects on image quality. In this work we
consider displacement, smear, and jitter as defined in \cite{pittelkau16}.


In this work, the model of the coupled bus and payload pointing control system is a continuous-time, linear time-invariant system. The output of the control system is a two-dimensional image motion on a focal plane. Since we assume the optical system is linear shift-invariant, image motion is related to the angular motion of an optical line of sight. Some (improper) requirements may define the output to be three-axis angular motion of a specified reference frame or two-axis angular motion of a vector direction (pointing). Any of these definitions can be accommodated. The image motion is obtained through a linear optical model (LOM), which is a matrix that maps structural and rigid-body modes to image motion. The image motion is used to compute the image motion MTF \cite{lucke92}. The total disturbance input is typically white noise or broadband noise plus spectral lines due to harmonic and subharmonic disturbances.
Broadband noise can be modeled as the output of a shaping filter driven by zero mean white noise. We assume that the system has reached a statistical steady-state so that the output is stationary. 

The covariance of the displacement, smear, and jitter and the mean smear vector (called pointing metrics in \cite{pittelkau16}) parameterize statistical image motion MTFs. The covariance matrices are computed from simulated pointing motion data, and it is possible to compute them from telemetry. The displacement, smear, and jitter covariances and the image motion MTFs can be used to verify pointing and image motion MTF requirements.

In \cite{pittelkau16}, the covariances are computed from the power spectral response of a dynamic system to disturbances comprising spectral lines, which are due to various vibration sources. Since the power spectrum of the system response comprises spectral lines, the integral of the frequency-weighted power spectrum becomes a summation of terms, and there is no approximation in the integration. However, this approach is approximate and inefficient when the power spectral density is continuous, since numerical integration such as rectangular or trapezoidal integration would be used, and is inaccurate when the system has lightly-damped structural modes.

The contribution of this Note is a method to efficiently and accurately compute the displacement, smear, and jitter covariance for a combined bus and payload pointing control system driven by stationary white noise. Bayard \cite{bayard04} derived algorithms to compute displacement and jitter covariance, where the jitter motion includes smear motion. This work revises and extends the results in \cite{bayard04} to compute covariance matrices for displacement, smear, and jitter. The results of the present Note also lead, as a special case, to a more efficient method to compute the displacement covariance than the method in \cite{bayard04}. The algorithm derived in this work is provided in the publicly available Image Motion OTF and Pointing Performance Analysis Toolbox (IMOTF-PPA) \cite{Pointing2025}, which runs in Matlab and Octave. The the algorithm is implemented in the function \texttt{covss.m} and demonstrated in the example \texttt{ex\_covss.m}.

The present Note is structured as follows. In Section~\ref{sec:probform}, the system model is defined and expressions for displacement, smear, and jitter and their covariances are defined. In Section~\ref{sec:CompCov}, we define a Lyapunov differential equation (LDE) whose solution is the covariance for accuracy, displacement, and smear, from which the jitter covariance is computed. The finite-horizon solution is obtained by rewriting the LDE in block state-space form, which is then solved by using a matrix exponential. This step provides a more efficient and robust solution using well-established numerical algorithms as described in~\cite{vanloan78}. Three examples in Section~\ref{sec:ex} demonstrate the algorithm's efficacy.

\section{Problem Formulation}
\label{sec:probform}



\subsection{System Model}

The image motion $\ve{p}(t)$ is the output of an asymptotically stable linear time-invariant (LTI) system driven by a zero-mean white noise process $\ve{u}(t)$. 
The system is represented in state-space form as
\begin{align}\label{eq:PtLTI}
\begin{split}
 \ve{\dot{x}}(t) & = \ve{A} \, \ve{x}(t) + \ve{B} \, \ve{u}(t) \\
   \ve{p}(t) & = \ve{C} \, \ve{x}(t),
\end{split}
\end{align}
where $\ve{x}(t) \in \mathbb{R}^{n_x}$ is the state vector, $\ve{u}(t) \in \mathbb{R}^{n_u}$, and $\ve{p}(t) \in \mathbb{R}^{n_p}$. When $\ve{C}$ is a linear optical model which maps structural and rigid-body  modes to image motion, the output $\ve{p}(t)$ is image motion.
Broadband noise input with non-flat power spectral density can be modeled by augmenting the dynamics, Eq.~\eqref{eq:PtLTI}, with shaping filters driven by white noise. We assume that the system has operated long enough to have reached a statistical steady state at time $t = 0$ so that $\ve{x}(t)$ and $\ve{p}(t)$ are stationary random processes for $t \ge 0$.
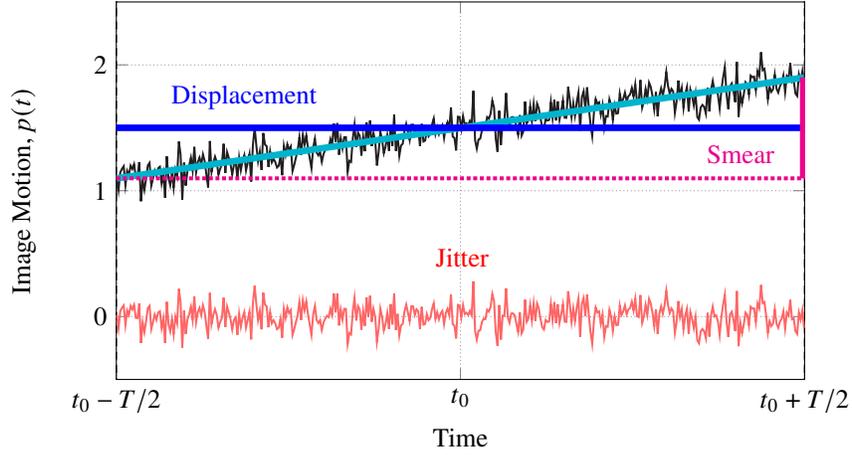
\begin{figure}[h!]
\centering
\begin{tikzpicture}
\definecolor{blue1}{RGB}{222,235,247}
\definecolor{blue2}{RGB}{158,202,225}
\definecolor{blue3}{RGB}{49,130,189}
%

\begin{axis}[ width = 0.65\columnwidth, 
            height = 0.4\columnwidth,
  	         grid=major, 
            grid style={densely dotted,white!60!black}, 
            ylabel= $\text{Image Motion,\,}p(t)$,
	           xlabel= $\text{Time}$,	
            xmin = 6.0, xmax = 14.0, ymin = -0.5, ymax = 2.5, 
            xtick={6,10,14},
            xticklabels={$t_0-T/2$,$t_0$,$t_0+T/2$},]

\addplot[-, black, densely dashed, line width = 0.5]coordinates{(6,-10)(6,10)};\label{pl:req}
\addplot[-, black, densely dashed, line width = 0.5]coordinates{(14,-10)(14,10)};\label{pl:req}

\addplot[Black, line width = 0.75] table[x expr = \thisrowno{0} ,y expr = \thisrowno{1} ,col sep=comma] {ImageMotion24.csv};\label{pl:Nom1}
\addplot[Turquoise, line width = 2.5] table[x expr = \thisrowno{0} ,y expr = \thisrowno{2} ,col sep=comma] {ImageMotion.csv};\label{pl:Nom1}
\addplot[red!60, line width = 0.75] table[x expr = \thisrowno{0} ,y expr = \thisrowno{1} ,col sep=comma] {ImageMotion.csv};\label{pl:Nom1}

\addplot[-, blue, solid, line width = 2.5]coordinates{(6,1.5)(14,1.5)};\label{pl:req}

\addplot[-, magenta, solid, line width = 3.5]coordinates{(14,1.0983)(14,1.9008)};\label{pl:req}

\addplot[-, magenta, densely dotted, line width = 1.5]coordinates{(6,1.0983)(14,1.0983)};\label{pl:req}


\end{axis}

\node[minimum width=1.3cm, minimum height=1.3cm, xshift = 4.6cm, yshift = +1.625cm](Cap) at (0,0) {\textcolor{red}{Jitter}};
\node[minimum width=1.3cm, minimum height=1.3cm, xshift = 1.70cm, yshift = +3.75cm](Cap) at (0,0) {\textcolor{blue}{Displacement}};

\node[minimum width=1.3cm, minimum height=1.3cm, xshift = 8.3cm, yshift = +3.0cm, rotate=0](Cap) at (0,0) {\textcolor{magenta}{Smear}};

\end{tikzpicture}
\vspace{-10pt}
\caption{Image motion (black) decomposed into displacement , 
    smear , and jitter.}
    \label{fig:ImageMotion}
\vspace{-30pt}
\end{figure}

\subsection{Displacement, Smear, and Jitter}
An image is integrated over an exposure interval of duration $T$ centered at time $t_0$. Figure~\ref{fig:ImageMotion} illustrates one-dimensional image motion ($n_p = 1$) over an exposure interval with the motion decomposed into displacement, smear, and jitter. The displacement $\ve{\bar{p}(}t_0, T)$ and smear vector $\ve{\bar{s}}(t_0, T)$ are the mean position and change in position of the image of a point source. The smear vector is $\ve{\bar{s}}(t_0, T) = T \ve{\bar{v}}(t_0, T)$ where the smear rate $\ve{\bar{v}}(t_0, T)$ is the average rate over the exposure interval. The displacement, smear rate, and smear depend on the centroid time $t_0$ and exposure duration $T$. The displacement and smear rate are defined according to \cite{pittelkau16} as follows
\begin{align}
\label{eq:disp}
\ve{\bar{p}}(t_0,T) & = \frac{1}{T} \int_{{-T}/{2}}^{{T}/{2}} \!\!\ve{p}(t_0+\alpha) \, \d\alpha, \\
\label{eq:srate}
\ve{\bar{v}}(t_0,T) & = \frac{12}{T^3} \int_{-{T}/{2}}^{{T}/{2}} \!\!\alpha 
   \ve{p}(t_0+\alpha) \, \d\alpha.
\end{align}
The formulae for displacement and smear rate provide the best fit linear trend
$\ve{\bar{p}}(t_0,T) + (t-t_0)\ve{\bar{v}}(t_0,T)$. In other
  words, $\ve{\bar{p}}(t_0,T)$ and $\ve{\bar{v}}(t_0,T)$ solve
\begin{align*}
  \min_{\bar{p},\bar{v}} \int_{-{T}/{2}}^{{T}/{2}} 
     \!\| \ve{p}(t_0+\alpha) - \ve{\bar{p}} - \ve{\bar{v}} \, \alpha \|^2 \, \d\alpha.
\end{align*}
 See \cite[\S2]{pittelkau16} for details. The instantaneous jitter $\ve{\psi}(t)$ for $t\in [t_0-T/2\,,\> t_0+T/2]$ is the residual motion after removing the displacement and smear from the pointing motion
\begin{align}
\label{eq:jitter}
\ve{\psi}(t) & = \ve{p}(t) 
   - \ve{\bar{p}}(t_0,T) - (t-t_0)\ve{\bar{v}}(t_0,T).
\end{align}

Since $\ve{p}(t)$ is stationary, we can shift the time reference $t_0$ to the start of the exposure, i.e., $t_0 = T/2$, and drop the time reference $t_0$ to simplify the derivation and notation. Thus, the displacement, smear, and jitter can be re-written as
\begin{align}
\label{eq:disp2}
\ve{\bar{p}}(T) & 
   = \frac{1}{T} \int_{0}^{T} \!\!\!\ve{p}(\alpha) \, \d\alpha, \\
\label{eq:srate2}
\ve{\bar{s}}(T)&= T\ve{\bar{v}}(T) = \frac{12}{T^2} \int_{0}^{T} \!\left(\alpha -\frac{T}{2}\right)
 \ve{p}(\alpha)\, \d\alpha, \\
\label{eq:jitter2}
\ve{\psi}(t) & = \ve{p}(t) 
   - \ve{\bar{p}}(T) - \left(t-\frac{T}{2} \right)\ve{\bar{v}}(T) .
\end{align}

\subsection{Double Integrator Form}

We now write the smear vector in terms of the states of a double integrator with the pointing motion $\ve{p}(t)$ as input
\begin{align}\label{eq:di}
\begin{split}
  \ve{\dot{z}}_1(t) & = \ve{p}(t) \\
  \ve{\dot{z}}_2(t) & = \ve{z}_1(t),
  \end{split}
\end{align}
with initial conditions $\ve{z}_1(0) = \ve{0}$ and $\ve{z}_2(0) = \ve{0}$. Integrate $\ve{\dot z}_1(t)$ from $t = 0$ to $t = T$ to get
\begin{align}
  \ve{z}_1(T) =\int_0^T \!\!\!\ve{p}(\alpha) \, \d\alpha.
\end{align}
Thus the displacement vector is $\ve{\bar p}(T) = \ve{z}_1(T)/T$. Integrate $\ve{\dot z}_2(t)$ to get
\begin{equation}
\begin{split}
  \ve{z}_2(T) &=\int_0^T \!\!\!\ve{z}_1(\tau) \, \d\tau \\ 
         &= \int_0^T \left[ \int_0^\tau \ve{p}(\alpha) d\alpha \right]\, \d\tau.
\end{split}
\end{equation}
Now, reverse the order of integration to obtain
\begin{equation}
        \begin{split}
       \ve{z}_2(T) &= \int_0^T \left[ \int_\alpha^T \ve{p}(\alpha) \d\tau \right]\, \d\alpha \\
         &= \int_0^T \!\!(T-\alpha)\ve{p}(\alpha) \, \d\alpha.
         \end{split}
\end{equation}
Then, from Eqs.~\eqref{eq:disp2} and~\eqref{eq:srate2}, the smear vector can be expressed in terms of $\ve{z}_1$ and $\ve{z}_2$
\begin{equation}
\begin{split}
   \ve{\bar s}(T) 
   &= \dfrac{6}{T} \int_0^T \!\!\!\ve{p}(\alpha) \, \d\alpha 
   - \dfrac{12}{T^2} \int_0^T \!\!(T - \alpha)\, \ve{p}(\alpha) \, \d\alpha \\
   &= \ve{L} \begin{bmatrix} \ve{z}_1 \\ \ve{z}_2 \end{bmatrix},
      \end{split}
\end{equation}
where 
\begin{equation}
   \ve{L} = \left[ \frac6T\,\ve{I}_{N_p} \quad {-}\frac{12}{T^2}\,\ve{I}_{N_p} \right].
\end{equation}

\subsection{Pointing Metrics (Covariances)}
 We have assumed that the state $\ve{x}(t)$ in Eq.~\eqref{eq:PtLTI} is stationary for $t \ge 0$ with mean $\E[\ve{x}(t)] = \ve{0}$ and covariance $\E[\ve{x}(t)\ve{x}(t)^\top] = \ve{P}$ ($\E$ is the statistical expectation operator). The symmetric matrix $\ve{P} \succeq 0$ is the solution to the algebraic Lyapunov equation \cite{anderson07} 
\begin{align}
  \label{eq:LyapEq}
  \ve{A} \ve{P} + \ve{P} \ve{A}^\top + \ve{B}\ve{B}^\top = \ve{0}.
\end{align}
The pointing motion $\ve{p}(t) = \ve{C}\ve{x}(t)$, $t \ge 0$, is zero mean with covariance
\begin{equation}
   \ve{\Sigma}_\text{A} = \E[\,\ve{p}(t)\ve{p}(t)^\top] = \ve{C}\ve{P}\ve{C}^\top.
\end{equation}
This is called the accuracy covariance or accuracy metric.
Additionally, $\ve{P}\succ 0$ when $(\ve{A},\,\ve{C})$ is controllable, but controllability is not a necessary assumption because we have assumed that the system is asymptotically stable.
The displacement covariances is
\begin{equation}\label{eq:combSD}
\begin{split}
   \ve{\Sigma}_\text{D} & =\E\!\left[\ve{\bar{p}}(T)\ve{\bar{p}}(T)^\top\right]\\
   &= \frac{1}{T^2} \E\!\left[\ve{z}_1(T)\ve{z}_1(T)^\top \right],
   \end{split}
\end{equation}
and the smear covariance is
\begin{equation}\label{eq:combSS}
\begin{split}
   \Sigma_\text{S} & =\E\!\left[\ve{\bar{s}}(T)\ve{\bar{s}}(T)^\top\right] \\ 
& = \ve{L} \, \E\!\left[\,\,\,\bmtx \ve{z}_1(T) \\ \ve{z}_2(T) \emtx 
\bmtx \ve{z}_1(T) \\ \ve{z}_2(T) \emtx^\top \right] \, \ve{L}^\top.
\end{split}
\end{equation}
The jitter covariance is, formally,
\begin{equation}
  \ve{\Sigma}_\text{J} = \E\!\left[\ve{\psi}(t)\ve{\psi}(t)^\top\right] 
\end{equation}
and can be readily computed:
it is shown in \cite[\S4,~App.~A]{pittelkau16} that the accuracy (total error) covariance $\ve{\Sigma}_\text{A}$ is
\begin{align}
\label{eq:acov}
\ve{\Sigma}_\text{A} = \ve{\Sigma}_\text{D} + \frac{1}{12}\ve{\Sigma}_\text{S} + \ve{\Sigma}_\text{J}.
\end{align}
so the jitter covariance is
\begin{align}
\label{eq:jcov}
\ve{\Sigma}_\text{J}  = \ve{\Sigma}_\text{A} - \ve{\Sigma}_\text{D}   - \frac{1}{12} \ve{\Sigma}_\text{S}.
\end{align}
Bayard \cite{bayard04} calls a result similar to Eq.~\eqref{eq:acov} the ``Conservation of Variance''. We call Eq.~\eqref{eq:acov} a balance equation.

\vspace{-10pt}
\section{Computation of the Covariances}\label{sec:CompCov}
A method to compute the displacement covariance using a Lyapunov differential equation (LDE) is given in \cite[Appendix]{bayard04}. We built upon Bayard's results to derive an augmented LDE to obtain the displacement and smear covariance from which the jitter covariance is computed. The augmented LDE is reformulated into a block state space form, $\ve{\dot{X}}(t)=\ve{M}\ve{X}(t)$, which is solved using a matrix exponential.
The matrix exponential solution is numerically faster and more accurate than directly integrating the LDE, especially in the presence of lightly damped structural modes. However, correctness of the algorithm and its implementation can be demonstrated
(though not necessarily proven) by integrating the LDE using trapezoidal integration for a problem with a smooth power spectral density, that is, without lightly-damped modes. Correctness can be demonstrated further by solving a simple problem analytically.

\vspace{-10pt}
\subsection{Augmented Lyapunov Differential Equation}

Equations~\eqref{eq:combSD} and \eqref{eq:combSS} define the displacement and smear covariances in terms of the covariance of $\ve{z_1}$ and $\ve{z_2}$. These results are applied by augmenting the pointing motion dynamics in Eq.~\eqref{eq:PtLTI} with the double integrator dynamics in Eq.~\eqref{eq:di} to get
\begin{align}
\label{eq:AugDefs}
\ve{\tilde{x}} = \bmtx \ve{x} \\ \ve{z}_1 \\ \ve{z}_2 \emtx, \,\,
\ve{\tilde{A}} = \bmtx \ve{A} & \ve{0} & \ve{0} \\ \ve{C} & \ve{0} & \ve{0} \\ \ve{0} & \ve{I}_{n_p} & \ve{0} \emtx, \,\,
\tilde{B} = \bmtx \ve{B} \\ \ve{0} \\ \ve{0} \emtx.
\end{align}
The covariance $\ve{\tilde P}(t) = \mathbb[\ve{\tilde x}(t)\ve{\tilde x}(t)^\top]$ of the augmented state is obtained by integrating the augmented Lyapunov differential equation~\cite{anderson07}
\begin{align}
\label{eq:LDE}
\begin{split}
&  \ve{\dot{\tilde{P}}}(t) = \ve{\tilde{A}} \ve{\tilde{P}}(t) + \ve{\tilde{P}}(t) \ve{\tilde{A}}^\top 
   + \ve{\tilde{B}}  \ve{\tilde{B}}^\top 
\end{split}
\end{align}
from $t = 0$ to $t = T$ with the initial covariance
\begin{align}
  \ve{\tilde{P}}(0)=
  \E[\ve{\tilde{x}}(0) \ve{\tilde{x}}(0)^\top] 
    = \bmtx \ve{P} & \ve{0} & \ve{0} \\ \ve{0} & \ve{0} & \ve{0} \\ \ve{0} & \ve{0} & \ve{0} \emtx.
\end{align}
 Partition $\ve{\tilde{P}}(t)$ conformally with the augmented state,
\begin{align}
\label{eq:PtilBlocks}
  \ve{\tilde{P}}(T) 
=\bmtx \ve{\tilde{P}}_{xx}(T) & \ve{\tilde{P}}_{xz_1}(T) & \ve{\tilde{P}}_{xz_2}(T) \\
\ve{\tilde{P}}_{z_1 x}(T) & \ve{\tilde{P}}_{z_1 z_1}(T) & \ve{\tilde{P}}_{z_1 z_2}(T) \\
\ve{\tilde{P}}_{z_2 x}(T) & \ve{\tilde{P}}_{z_2 z_1}(T) & \ve{\tilde{P}}_{z_2 z_2}(T).
\emtx.
\end{align}
The displacement and smear covariances are given by
\begin{align}
\label{eq:pcov}
\ve{\Sigma}_\text{D}
   & = \frac{1}{T^2} \ve{\tilde{P}}_{z_1 z_1}(T), \\
\label{eq:scov}
\ve{\Sigma}_\text{S}
   & = T^2 \ve{L} \bmtx \ve{\tilde{P}}_{z_1 z_1}(T) & \ve{\tilde{P}}_{z_1 z_2}(T) \\
               \ve{\tilde{P}}_{z_2 z_1}(T) & \ve{\tilde{P}}_{z_2 z_2}(T) \emtx \ve{L}^\top. 
\end{align}
Equations~\eqref{eq:pcov} and~\eqref{eq:scov} follow from Eqs.~\eqref{eq:combSD} and~\eqref{eq:combSS}. The jitter covariance $\ve{\Sigma}_\text{J}$ is computed from Eq~\eqref{eq:jcov}. In the next section, we derive a block state space form of~\eqref{eq:LDE}.

\subsection{Block State Space Form}

The LDE \eqref{eq:LDE} can be reformulated into the block state space form $\ve{\dot{X}}(t)=\ve{M}\ve{X}(t)$. Substitute the augmented state matrices
$(\ve{\tilde{A}},\ve{\tilde{B}})$ defined in Eq.~\eqref{eq:AugDefs} into the
Lyapunov differential equation \eqref{eq:LDE}. Partitioning the
covariance matrix $\ve{\tilde{P}}(t)$ as done in
Eq.~\eqref{eq:PtilBlocks} yields
\begingroup
\allowdisplaybreaks
\begin{align}
\label{eq:Pxx}
\ve{\dot{\tilde{P}}}_{xx}(t) & = \ve{A} \ve{\tilde{P}}_{xx}(t)+ \ve{\tilde{P}}_{xx}(t) \ve{A}^\top 
   + \ve{B}\ve{B}^\top, \\
\label{eq:Pxz1}
\ve{\dot{\tilde{P}}}_{xz_1}(t) & = \ve{A} \ve{\tilde{P}}_{xz_1}(t)+ \ve{\tilde{P}}_{xx}(t) \ve{C}^\top,
\\
\label{eq:Pxz2}
\ve{\dot{\tilde{P}}}_{xz_2}(t) & = \ve{A} \ve{\tilde{P}}_{xz_2}(t)+ \ve{\tilde{P}}_{xz_1}(t),
\\
\label{eq:Pz1z1}
\ve{\dot{\tilde{P}}}_{z_1z_1}(t) & = \ve{C} \ve{\tilde{P}}_{xz_1}(t)
+ \ve{\tilde{P}}_{xz_1}(t)^\top \ve{C}^\top,
\\
\label{eq:Pz1z2}
\ve{\dot{\tilde{P}}}_{z_1z_2}(t) & = \ve{C} \ve{\tilde{P}}_{xz_2}{(t)}+ \ve{\tilde{P}}_{z_1z_1}(t),
\\
\label{eq:Pz2z2}
\ve{\dot{\tilde{P}}}_{z_2z_2}(t) & = \ve{\tilde{P}}_{z_1z_2}(t)+ \ve{\tilde{P}}_{z_1z_2}(t)^\top.
\end{align}
\endgroup
The initial condition for Eq.~\eqref{eq:Pxx} is $\ve{\tilde{P}}_{xx}\ve{(0)} = \ve{P}$ where $\ve{P}$ solves the Lyapunov algebraic equation~\eqref{eq:LyapEq}. Hence $\ve{\dot{\tilde P}}_{xx}(t) = \ve{P}$ for $t \ge 0$ is the unique solution to the differential equation~\eqref{eq:Pxx}.
The remaining differential equations \eqref{eq:Pxz1}--\eqref{eq:Pz2z2} have zero initial conditions.
These equations cannot be grouped together into the block from $\ve{\dot{X}}(t) = \ve{M} \ve{X}(t)$.
To address this issue, decompose the solution of Eq. \eqref{eq:Pz1z1} as
\begin{align}
\label{eq:Pz1z1Aux}
 \ve{\tilde{P}}_{z_1z_1}(t) & = \ve{Y}_{z_1z_1}(t) + \ve{Y}_{z_1z_1}(t)^\top,
\end{align}
where the auxiliary variable is
\begin{align*}
  \ve{\dot{Y}}_{z_1z_1} & = \ve{C} \ve{\tilde{P}}_{xz_1}(t),   
  \,\,  \mbox{ with } \,\,
  \ve{Y}_{z_1z_1}(0) =\ve{0}.
\end{align*}
is in block state-space form.
We can similarly write Eqs.~\eqref{eq:Pz1z2}
and~\eqref{eq:Pz2z2} as
\begin{align}
\label{eq:Pz1z2Aux}
 \ve{\tilde{P}}_{z_1z_2}(t)&  =\ve{Y}_{z_1z_2}(t)+\ve{Y}_{z_1z_2}(t)^\top + \ve{W}_{z_1z_2}(t),
 \\
\label{eq:Pz2z2Aux}
 \ve{\tilde{P}}_{z_2z_2}(t)&  = \ve{Y}_{z_2z_2}(t)+\ve{Y}_{z_2z_2}(t)^\top.
\end{align}
\noindent where the auxiliary variables are
\begin{align}
\label{eq:Aux1}
  \ve{\dot{Y}}_{z_1z_2}(t) & = \ve{Y}_{z_1z_1}(t), \\
  \label{eq:Aux2}
  \ve{\dot{W}}_{z_1z_2}(t) & = \ve{C}\ve{\tilde{P}}_{xz_2}(t), \\
  \label{eq:Aux3}
  \ve{\dot{Y}}_{z_2z_2}(t) & = \ve{W}_{z_1z_2}(t)+2 \ve{Y}_{z_1z_2}(t)
\end{align}
with initial conditions $\ve{Y}_{z_1z_2} = \ve{W}_{z_1z_2} = \ve{Y}_{z_2z_2} = \ve{0}$.

The various differential equations involving blocks of $\ve{\tilde P}$ and auxiliary variables are assembled into the block state space form
\begin{align}
\label{eq:XdotMX}
\footnotesize
\underbrace{\frac{d}{dt}
\left[ \begin{array}{c}
\ve{Y}_{z_2z_2} \\ \ve{W}_{z_1z_2} \\ \ve{\tilde{P}}_{x z_2} \\ \ve{Y_{z_1z_2}} \\ 
\hline \ve{Y}_{z_1z_1} \\ \ve{\tilde{P}}_{xz_1} \\ \ve{F} 
\end{array}\right]}_{\ve{\dot{X}}}
\!\!=\!\!
\underbrace{\left[ \begin{array}{cccc|ccc} 
\ve{0} \!\!\!&\!\!\! \ve{I}_{n_p} \!&\!\!\! \ve{0} \!\!\!&\!\!\! 2\ve{I}_{n_p} \!&\!\!\! \ve{0} \!&\!\!\! \ve{0} \!\!\!&\!\!\! \ve{0} \\
\ve{0} \!\!\!&\!\!\! \ve{0} \!\!\!&\!\!\! \ve{C} \!\!\!&\! \ve{0} \!\!\!&\!\!\! \ve{0} \!\!\!&\!\!\! \ve{0} \!\!\!&\!\!\! \ve{0} \\
\ve{0} \!\!\!&\!\!\! \ve{0} \!\!\!&\!\!\! \ve{A} \!\!\!&\! \ve{0} \!\!\!&\!\!\! \ve{0} \!\!\!&\!\!\! \ve{I}_{n_x} \!&\! \ve{0} \\
\ve{0} \!\!\!&\!\!\! \ve{0} \!\!\!&\!\!\! \ve{0} \!\!\!&\! \ve{0} \!\!\!&\!\!\! \ve{I}_{n_p} \!\!\!&\!\!\! \ve{0} \!\!\!&\!\!\! \ve{0} \\ \hline
\ve{0} \!\!\!&\!\!\! \ve{0} \!\!\!&\!\!\! \ve{0} \!\!\!&\! \ve{0} \!\!\!&\!\!\! \ve{0} \!\!\!& \ve{C} \!\!\!&\!\!\! \ve{0} \\
\ve{0} \!\!\!&\!\!\! \ve{0} \!\!\!&\!\!\! \ve{0} \!\!\!&\! \ve{0} \!\!\!&\!\!\! \ve{0} \!\!\!& \ve{A} \!\!\!&\!\!\! \ve{P}\ve{C}^\top \\
\ve{0} \!\!\!&\!\!\! \ve{0} \!\!\!&\!\!\! \ve{0} \!\!\!&\! \ve{0} \!\!\!&\!\!\! \ve{0} \!\!\!& \ve{0} \!\!\!&\!\!\! \ve{0} \\
\end{array} \right]}_{\ve{M}} 
\underbrace{\left[ \begin{array}{c}
\ve{Y}_{z_2z_2} \\ \ve{W}_{z_1z_2} \\ \ve{\tilde{P}}_{x z_2} \\ \ve{Y}_{z_1z_2} \\ 
\hline \ve{Y}_{z_1z_1} \\ \ve{\tilde{P}}_{xz_1} \\ \ve{F} 
\end{array}\right]}_{\ve{X}},
\end{align}
where the lower right block independently describes the displacement covariance.
The initial conditions are $\ve{F}(0)=\ve{I}_{n_p}$ and the
remaining block states have zero initial conditions. The last block row of this differential equation is $\ve{\dot{F}}(t)=\ve{0}$ so that $\ve{F}(t)=\ve{I}_{n_p}$
for all $t\ge 0$.  
The dimension of $\ve{M}$ is
$(5n_p+2n_x)\times (5n_p+2n_x)$.

\vspace{-10pt}
\subsection{Matrix Exponential Solution}\label{ss:MES}
The solution to the matrix differential equation $\ve{\dot{\tilde X}}(t) = \ve{M}\ve{X}(t)$ at time $t = T$ is $\ve{X}(T) = e^{\ve{M}}T\ve{X}(0)$, where the initial condition is $\ve{X}(0) = [\,\ve{0} ~~ \ve{I}_{n_p}]^\top$  (see \cite{vanloan78} for details on this basic result). Then $\ve{X}(T)$ is simply the last $n_p$ columns of the matrix exponential $e^{\ve{M}T}$,
\begin{align}
\label{eq:expMTX0}
 \ve{X}(T)=e^{\ve{M}T} \ve{X}(0) = \left[\begin{array}{c}
\ve{Y}_{z_2z_2}(t) \\ \ve{W}_{z_1z_2}(t) \\ \ve{\tilde{P}}_{x z_2}(t) \\ \ve{Y}_{z_1z_2}(t) \\ 
\hline \ve{Y}_{z_1z_1}(t) \\ \ve{\tilde{P}}_{xz_1}(t) \\ \ve{I}_{n_p}
\end{array}\right].
\end{align}

The augmented covariance $\ve{\tilde P}(t)$ is computed from the solution $\ve{X}(T)$. Since $\ve{\tilde{P}}_{xx}(t) = \ve{P}$ for $t \ge 0$, we have that $\ve{\tilde{P}}_{xx}(t) = \ve{P}$, and $\ve{\tilde{P}}_{z_1z_1}(t)$, $\ve{\tilde{P}}_{z_1z_2}(t)$, and $\ve{\tilde{P}}_{z_2z_2}(t)$ are computed from Eqs.~\eqref{eq:Pz1z1Aux}--\eqref{eq:Pz2z2Aux} and the auxiliary variables defined in Eqs.~\eqref{eq:Aux1}--\eqref{eq:Aux3}. The displacement, smear, and jitter covariances are computed from $\ve{\tilde P}(t)$ using Eqs.~\eqref{eq:pcov}, \eqref{eq:scov}, and~\eqref{eq:jcov}, respectively.

The computation can be simplified if only the displacement covariance is required, but not the smear covariance. The auxiliary variable $\ve{Y}_{z_1z_1}$ is obtained using a smaller
differential equation obtained from the lower right block of $\ve{M}$

\begin{align}
\label{eq:XdotMXsmall}
\frac{d}{dt}
\left[ \begin{array}{c}
\ve{Y}_{z_1z_1} \\ \ve{\tilde{P}}_{xz_1} \\ \ve{F} 
\end{array}\right]
=
\left[
 \begin{array}{ccc} 
  \ve{0} & \ve{C} & \ve{0} \\
  \ve{0} & \ve{A} & \ve{P}\ve{C}^\top \\
  \ve{0} & \ve{0} & \ve{0} \\
 \end{array} \right]
\left[
\begin{array}{c}
\ve{Y}_{z_1z_1} \\ \ve{\tilde{P}}_{xz_1} \\ \ve{F} 
\end{array}\right].
\end{align}
The initial conditions are $\ve{Y}_{z_1z_1}(0)=\ve{0}$, $\ve{\tilde{P}}_{xz_1}(0)=\ve{0}$,
$\ve{F}(0)=\ve{I}_{n_p}$.  As described above, the solution $\ve{Y}_{z_1z_1}(t)$ can
obtained using a matrix exponential and the auxiliary variable.  This
can then be used to construct $\ve{\tilde{P}}_{z_1z_1}(t)$ and hence the
displacement covariance $\ve{\Sigma}_D=(1/T^2) \ve{\tilde{P}}_{z_1z_1}(t)$
by Eq.~\eqref{eq:pcov}. It follows from Eq.~\eqref{eq:acov}
  that $\ve{\Sigma}_\text{A}-\ve{\Sigma}_\text{D}=\ve{\Sigma}_\text{J}+\frac{1}{12}\ve{\Sigma}_\text{S}$. Thus
  $\ve{\Sigma}_\text{A}-\ve{\Sigma}_\text{D}$ corresponds to the sum of jitter and (scaled)
  smear covariances. This quantity was previously called ``jitter'' in
  \cite{bayard04} and elsewhere. It has been more recently renamed
  ``smitter'' in \cite{pittelkau16}.  Combining the effects into
  ``smitter'' is less useful. It is preferred to separate smear and
  jitter due to their different effects on image quality.  Thus
computing displacement covariance on its own requires solving the
differential equation \eqref{eq:XdotMXsmall} with smaller dimension
$(2n_p+n_x)\times (2n_p+n_x)$.   

There is a similar approach in \cite{bayard04} but with a matrix exponential of dimension $3n_x \times 3n_x$. It is typical that $n_x\gg n_p$. As a result, the matrix exponential obtained from Eq. \ref{eq:XdotMXsmall} is computationally less costly than using the matrix given in \cite{bayard04}.  Moreover, the matrix exponential in Eq. \ref{eq:XdotMX} has dimension
$(5n_p + 2 n_x) \times (5 n_p + 2 n_x)$ and can be used to compute the smear, displacement, and jitter covariances.  Thus Eq. \ref{eq:XdotMX} provides more information and, if $n_x \gg n_p$, is computationally less costly than using the matrix given in \cite{bayard04}.

\vspace{-10pt}
\section{Numerical Examples}
The displacement, smear, and jitter covariances
are computed for three examples. The first example is a first order process from reference \cite{bayard04}. The second example is a pointing process generated from a MIMO system. The third example is a satellite with flexible appendages. Other examples are included in the IMOTF-PPA Toolbox~\cite{Pointing2025}.

\label{sec:ex}
\subsection{First-Order Pointing Process}

This example is from Section IV.A of \cite{bayard04}. The first-order state and output equations are
\begin{align}
\begin{split}
 \dot{x}(t) & = a \, x(t) + b \, u(t) \\
   p(t) & =   x(t).
\end{split}
\end{align}
The output gain is set to $c=1$, the input gain $b=\sqrt{q}$ (where $q$ is defined in~\cite{bayard04}), and we assume that $a<0$ so that the process is stable. The sign of $a$ is flipped compared to \cite{bayard04} to keep the notation consistent with the derivations in the previous sections. Here, we compute the displacement, smear, and jitter covariances. We can symbolically solve the algebraic Lyapunov
equation~\eqref{eq:LDE} and the blocks of the Lyapunov differential
equation \eqref{eq:Pxx}--\eqref{eq:Pz2z2}. The solution of the
algebraic Lyapunov equation is the accuracy covariance
\begin{equation}
\Sigma_\text{A} = P=-\frac{b^2}{2a}.    
\end{equation}
Substituting the solutions of the Lyapunov differential equation into Eqs.~\eqref{eq:jcov}--\eqref{eq:pcov} yields the following expressions
\begin{align}
\label{eq:CdispSym}
\Sigma_\text{D}
& = \frac{2P}{(aT)^2} \left[ e^{aT} - 1 - aT \right] \\
\label{eq:CsmearSym}
\Sigma_\text{S}
& =  \frac{24P}{(aT)^4} \left[ 12( aTe^{aT}+1-e^{aT}) \right. \\
\nonumber
& \qquad\qquad \left. -3(aT)^2(1+e^{aT})- (aT)^3  \right] \\
\label{eq:CjitterSym}
\Sigma_\text{J}
& =  \frac{P}{(aT)^4} \left[ 24( e^{aT}-1-aTe^{aT}) \right. \\
\nonumber
& \qquad\qquad \left. + 4(aT)^2(2+e^{aT})
+ 4(aT)^3 + (aT)^4 \right].     
\end{align}

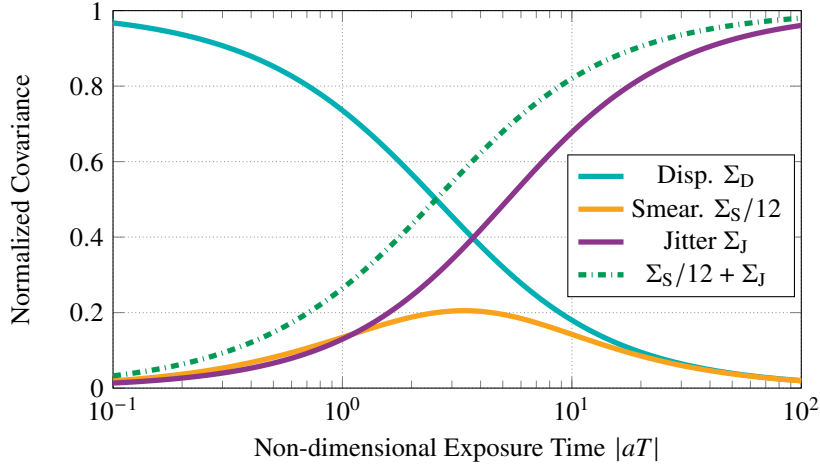
\begin{figure}[h!]
\centering
\begin{tikzpicture}
\definecolor{blue1}{RGB}{222,235,247}
\definecolor{blue2}{RGB}{158,202,225}
\definecolor{blue3}{RGB}{49,130,189}
%

\begin{groupplot}[group style={
                      	group name=myplot,
                      	group size= 1 by 1,
                        vertical sep=2.25cm,
                        horizontal sep = 1.75cm},
                      	height=0.4\columnwidth,
                      	width = 0.65\columnwidth,
                      	xmajorgrids=true,
			ymajorgrids=true,
			 grid style={densely dotted,white!60!black},
			  xmin = 0.1, xmax = 100,
			  ymin = 0, ymax = 1,
              xmode=log,
              legend style={at={(+0.66,0.43)},anchor=west},
			   ]]

\nextgroupplot[	  ylabel= $\text{Normalized Covariance}$,
				 xlabel= $\text{Non-dimensional Exposure Time} \,\abs{aT}$,
				 ]
\addplot[TealBlue, line width = 2.0] table[x expr = \thisrowno{0} ,y expr = \thisrowno{1} ,col sep=comma] {FirstOrder.csv};\label{pl:Nom1}
\addplot[YellowOrange, line width = 2.0] table[x expr = \thisrowno{0} ,y expr = \thisrowno{2} ,col sep=comma] {FirstOrder.csv};\label{pl:Nom2}
\addplot[Fuchsia, line width = 2.0] table[x expr = \thisrowno{0} ,y expr = \thisrowno{3} ,col sep=comma] {FirstOrder.csv};\label{pl:Nom3}
\addplot[ForestGreen, dashdotted,line width = 2.0] table[x expr = \thisrowno{0} ,y expr = \thisrowno{4} ,col sep=comma] {FirstOrder.csv};\label{pl:Nom4}


\addlegendentry{Disp. $\Sigma_\text{D}$}
\addlegendentry{Smear. $\Sigma_\text{S}/12$}
\addlegendentry{Jitter $\Sigma_\text{J}$}
\addlegendentry{$\Sigma_\text{S}/12+\Sigma_\text{J}$}

\end{groupplot}
\end{tikzpicture}
\caption{Normalized pointing covariances for a first-order process
    as a function of non-dimensional exposure time $\abs{aT}$.}
    \label{fig:FirstOrderExample}
    \vspace{-10pt}
\end{figure}

It can be verified that these covariances satisfy the balance equation, Eq.~\eqref{eq:acov}. Equation \eqref{eq:CdispSym} agrees with
the expression for the displacement covariance in Section IV.A of~\cite{bayard04}. The smear covariance Eq.~\eqref{eq:CsmearSym} is new,
and the jitter covariance, Eq.~\eqref{eq:CjitterSym}, is different from~\cite{bayard04} due to our definition of jitter. However, from Eqs.~\eqref{eq:CsmearSym} and~\eqref{eq:CjitterSym}, $\Sigma_\text{S}/12 + \Sigma_\text{J}$ yields the jitter (“smitter”) covariance in~\cite{bayard04}.
The parameters $a$ and $T$ appear in Eqs.~\eqref{eq:CdispSym}--\eqref{eq:CjitterSym} as the non-dimensional product $aT$. The input gain $b$
appears only in $P$, so the normalized covariances $\Sigma_\text{D}/P$, $\Sigma_\text{S}/P$, $\Sigma_\text{J}/P$ do not depend on $b$. Figure~\ref{fig:FirstOrderExample} shows
the normalized covariances as functions of $\abs{aT}$. The jitter covariance increases with increasing exposure
time $T$. The smear covariance is small at short and long exposure times, and is largest at $\abs{aT} \simeq 3$ where the $\Sigma_\text{D}$ and $\Sigma_\text{J}$ curves cross. The dashed curve is the “smitter”, $\Sigma_\text{S}/12 + \Sigma_\text{J}$, which is the jitter curve in Figure 4 of~\cite{bayard04}. The jitter covariance defined in this work is always smaller than the “smitter” covariance.

\subsection{Multi-Input Multi-Output Example}

Consider the following two-input, two-output
system
\begin{align}\label{eq:MIMO}
\ve{G}(s) = \bmtx
\DS\frac{100}{s^2 + 1.4 s + 100} &
\DS\frac{45}{s^2 + 30 s + 225} \\
\DS\frac{12.8}{s^2 + 16 s + 64} &
\DS\frac{225}{s^2 + 1.2 s + 225}
\emtx.
\end{align}
Equation~\eqref{eq:MIMO}~does not represent an actual system, but is designed to demonstrate the covariance algorithm. The individual entries $(i,j)$ of this transfer function matrix are second-order systems. Table~\ref{tab:MIMO} provides the numerical values of their DC-gains $k_{ij}$, natural frequencies $\omega_{\text{n}_{ij}}$, and damping ratios $\zeta_{ij}$.
\begin{table}[ht!]
\caption{DC-gains $k_{ij}$, natural frequencies $\omega_{\text{n}_{ij}}$, and damping ratios $\zeta_{ij}$ characterizing $\ve{G}(s)$}\label{tab:MIMO}
\vspace{-20pt}
\begin{center}
\begin{tabular}
{ P{1.7cm} P{1.0cm} P{1.0cm}  P{1.0cm}}
  Entry $(i,j)$ & $k_{ij}$ & $\omega_{\text{n}_{ij}}$ & $\zeta_{ij}$\\
\hline
$(1,1)$ & $1$ & $10$ & $0.07$\\
$(1,2)$ & $0.2$ & $15$ & $1$ \\
$(2,1)$ & $0.2$ & $8$ & $1$\\
$(2,2)$ & $1$ & $15$ & $0.04$\\
\end{tabular}
\end{center}
\end{table}
The system $\ve{G}(s)$ has an $8$th-order minimal state-space realization $(\ve{A},\ve{B},\ve{C})$.
The realization is not unique but all minimal realizations have the
same input/output dynamics and hence yield the same pointing
covariances. The matrix exponential method in
Section~\ref{ss:MES} was used to compute the pointing
covariances for an exposure time $T=0.3\,$s. This yields
\begin{align*}
\ve{\Sigma}_\text{A} & = \bmtx  35.86 & 2.168 \\ 2.168 & 93.83 \emtx &
\ve{\Sigma}_\text{D} & = \bmtx  17.24 &  0.994 \\ 0.994 & 15.15 \emtx \\
\ve{\Sigma}_\text{S} & = \bmtx 184.5 & 9.252 \\ 9.252 &  596.7 \emtx&
\ve{\Sigma}_\text{J} & = \bmtx 3.244 & 0.403 \\ 0.403 & 28.96 \emtx.
\end{align*}
The covariances satisfy the balance equation, Eq.~\eqref{eq:acov}. The state and output dimensions are $n_x=8$
and $n_p=2$.  Thus the dimension of the matrix exponential is
$(5n_p+2n_x)=26$. 
\subsection{Satellite Example}
The third example demonstrates the pointing performance of a spacecraft with a rigid center body and two symmetrical flexible solar arrays. Each solar array has three flexible modes.
The linearized satellite dynamics are calculated using the Satellite Dynamics Toolbox \cite{alazard08, SDT} by following the instructions given in \cite{Demo1}. Only the rotational motion is considered in the example. The satellite's rotational dynamics $\ve{G}_\text{SC}$ are defined with respect to a Cartesian coordinate system $B$ fixed to the satellite center of gravity. The numerical state-space data is provided in the Appendix. Eighteen states fully describe the spacecraft dynamics: six states $\ve{x}_\text{r}$ for the rigid motion and twelve states $\ve{x}_\text{f}$ for the flexible modes. The attitude of the satellite in the body frame $B$ is defined by the angles $\Phi$, $\Theta$, $\Psi$ around the $x_B$, $y_B$, $z_B$ axes, respectively. These angles also define the output $\ve{y}_\text{SC}$ of the system, i.e., $\ve{y}_\text{SC}=\left[ \Phi, \, \Theta, \, \Psi\right]^\top$. 
The attitude of the spacecraft is measured by a star tracker, which is modeled as a first-order low-pass transfer function $\ve{G}_\text{STS}$ with a bandwidth of $50\,$rad/s.
The attitude control torque vector $\ve{\tau} = \left[ \tau_x, \tau_y, \tau_z\right]$ is effected by three body-axis aligned reaction wheels $\ve{G}_\text{RW}$ whose dynamics are second-order low-pass transfer functions with natural frequency of $628\,$rad/s and damping ratio of $0.7$.
The controller $C$ is composed of three SISO  proportional derivative (PD) controllers. The controller stabilizes the satellite and provides pointing capabilities about its three body axes. Here, a zero-reference command $\ve{r}=\left[0,0,0\right]^\top$ is considered.
The PD controller achieves tracking bandwidths $\omega_\text{C}$ of $0.7\,$rad/s ($\Phi$ channel), $0.4\,$rad/s ($\Theta$ channel), and $0.7\,$rad/s ($\Psi$ channel) with a phase margin of at least $55\,$deg on each channel.
The derivative term is implemented as $k_\text{D} s/(T_\text{K} s+1)$ with time constant $T_{\text{K}}$. This is a smoothed derivative and ensures that the PD controllers are implementable. The explicit controller values of the proportional gains $k_\text{P}$ and derivative gains $k_\text{D}$, as well as the time constants are given in Table~\ref{tab:Gains}.
\begin{table}[ht!]
\caption{Controller gains}\label{tab:Gains}
\begin{center}
\begin{tabular}
{ P{1.5cm} P{1.0cm} P{1.0cm}  P{1.0cm}}
  Channel & $k_\text{P}\,\left[ \frac{\text{Nm}}{\text{rad}}\right]$ & $k_\text{D}\,\left[ \frac{\text{Nm}}{\text{rad}}\right]$ & $T_\text{K}\,[\text{s}]$\\
\hline
$\Phi$ & $4.9$ & $741$ & $1.07$\\
$\Theta$ & $1.24$ & $329$ & $0.02$ \\
$\Psi$ & $0.576$ & $92.34$ & $0.76$\\
\end{tabular}
\end{center}
\end{table}

 \begin{figure}[h!]
    \centering
        \begin{tikzpicture}[blockdiag]
	
	
	\node[block](Plant){$\ve{G}_\text{SC}$}; 
	\node[block,left = of Plant, xshift = +0.0cm, yshift = +0.0cm](TVC) {$ \ve{G}_\text{RW}$}; 
	\node[block, left = of TVC](C){$\ve{C}$};
	\node[sum, left = of C, xshift = -0.0cm](Sum1){};
	\node[sum, right = of Plant, xshift=0.2cm](Sum2){};		
	\node[connector, right = of Sum2](Con1){};		
	\node[block, below = of Plant, yshift = -0.0cm](STS2){$\ve{G}_{\text{STS}}$}; 
    \node[block, above = of Sum2, yshift = +0.25cm](GF){$\ve{G}_{\text{F}_i}$}; 
    \node[block, right = of Con1, yshift = +0.0cm](Sel){$\bsmtx 1&0&0\\ 0&0&1 \esmtx$};



\draw[->](TVC.east) -- (Plant.west)node[pos=0.5, yshift =0.2cm] {$\ve{\tau}$};
\draw[->](C.east) -- (TVC.west);
\draw[->](Sum1.east) -- (C.west);
\draw[->]([yshift = -0.0cm, xshift =0.00cm]Plant.east) -- (Sum2.west)node[pos=0.5, yshift =0.2cm] {$\ve{y}_\text{SF}$};
\draw[-](Sum2.east) -- (Con1.west)node[pos=0.5, yshift =0.2cm] {$\ve{y}$};



\draw[<-]([yshift = +0.0cm, xshift = 0.00cm]GF.north)-- +(0.0cm, +0.8cm)node[above, name = e, xshift = 0.2cm, yshift = -0.1cm]{$\ve{d}$};

\draw[->]([yshift = +0.0cm, xshift = 0.00cm]Sel.east)-- +(0.5cm, +0.0cm)node[above, name = e, xshift = 0.125cm, yshift = -0.1cm]{$\ve{p}$};

\draw[->](Con1.south) |- (STS2.east);
\draw[->](STS2.west) -| (Sum1.south)node[left, yshift =-0.1cm, xshift = 0.5cm] {$-$};
\draw[->](GF.south) -| (Sum2.north)node[pos=0.0, yshift =-0.2cm, xshift = 0.25cm] {$\ve{y}_\text{F}$};
\draw[->](Con1.east) -- (Sel.west);


\draw[<-](Sum1.west) -- +(-0.5cm, 0.0cm)node[above, name = n, yshift = -0.025cm]{$\ve{r}=\ve{0}$};

\end{tikzpicture}

	
        \caption{Covariance analysis interconnection}
        \vspace{-10pt}
        \label{Fig:AnalysisConn}
\end{figure}
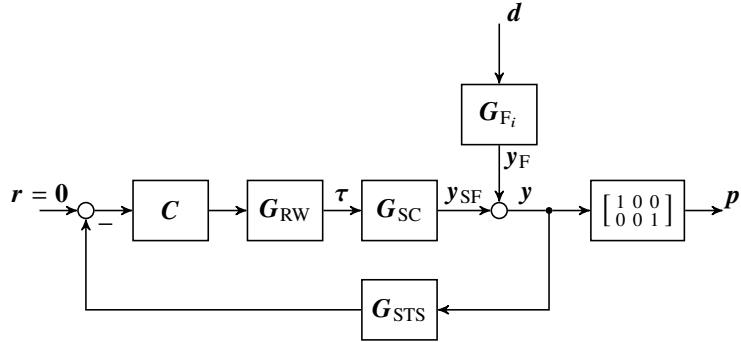
 Figure \ref{Fig:AnalysisConn} shows the closed loop interconnection used for the pointing performance analysis. The analysis only considers the pointing motion $\ve{p}$ about the $x$- and $z$-axis. The pointing motion is perturbed by a disturbance $\ve{y}_\text{F}$ acting at the output $\ve{y}_\text{SC}$, i.e., $\ve{p} = \bsmtx 1&0&0\\0&0&1\esmtx [\ve{y}_\text{SF}+\ve{y}_\text{F}]$. The signal $\ve{y} = \ve{y}_\text{SF}+\ve{y}_\text{F}$ is measured by the star tracker. 
 The disturbance is modeled by a shaping filter $\ve{G}_{\text{F}_i}$ 
 driven by a zero-mean white noise signal $\ve{d}$. 
The low-pass shaping filter defines the disturbance spectrum. We analyze the effect of three different parameters on the pointing covariances: exposure length T, tracking bandwidth $\omega_\text{C}$, and the disturbance bandwidth.
 
\begin{figure}[h!]
\centering
\begin{tikzpicture}
\definecolor{blue1}{RGB}{222,235,247}
\definecolor{blue2}{RGB}{158,202,225}
\definecolor{blue3}{RGB}{49,130,189}
%

\begin{groupplot}[group style={
                      	group name=myplot,
                      	group size= 2 by 1,
                        vertical sep=2.25cm,
                        horizontal sep = 1.75cm},
                      	height=0.4\columnwidth,
                      	width = 0.65\columnwidth,
                      	xmajorgrids=true,
			ymajorgrids=true,
			 grid style={densely dotted,white!60!black},
			  xmin = 0.0001, xmax = 10,
			  ymin = 0, ymax = 670,
              xmode=log,
              legend style={at={(0.975,0.45)},anchor=east},
			   ]]

\nextgroupplot[	
				  ylabel= $\text{Covariance}$,
				 xlabel= $\text{Exposure Time} \,T  \,{[\text{s}]}$,
				 ]
\addplot[blue3, line width = 2.0] table[x expr = \thisrowno{0} ,y expr = \thisrowno{1} ,col sep=comma] {CovXX.csv};\label{pl:Nom1}
\addplot[TealBlue, line width = 2.0] table[x expr = \thisrowno{0} ,y expr = \thisrowno{2} ,col sep=comma] {CovXX.csv};\label{pl:Nom2}
\addplot[YellowOrange, line width = 2.0] table[x expr = \thisrowno{0} ,y expr = \thisrowno{3} ,col sep=comma] {CovXX.csv};\label{pl:Nom3}
\addplot[Fuchsia, line width = 2.0] table[x expr = \thisrowno{0} ,y expr = \thisrowno{4} ,col sep=comma] {CovXX.csv};\label{pl:Nom4}

\addlegendentry{Acc. $\Sigma_\text{A}$}
\addlegendentry{Disp. $\Sigma_\text{D}$}
\addlegendentry{Smear $\Sigma_\text{S}/12$}
\addlegendentry{Jitter $\Sigma_\text{J}$}

\end{groupplot}
\end{tikzpicture}
\caption{Pointing covariances for increasing exposure times about x-axis (left) and z-axis (right).}
\label{fig:Exposure}
\end{figure}
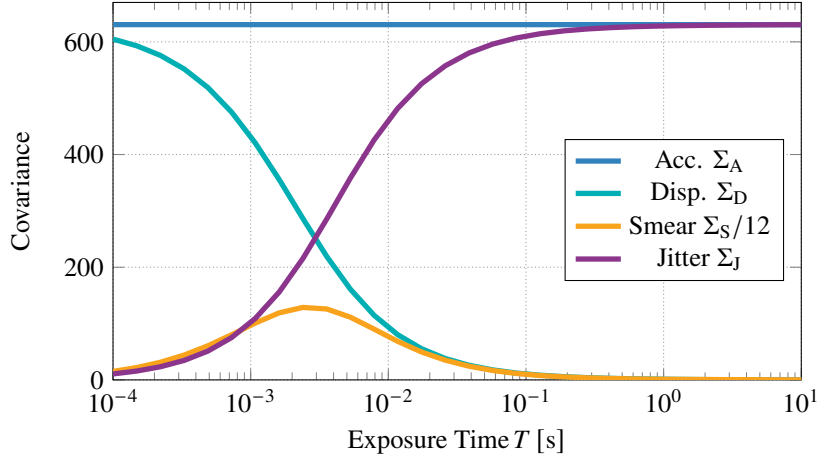

First, the pointing metrics for $30$ logarithmically spaced exposure times ranging from $T=0.1\,$ms to $10\,$s and a filter 
\begin{equation}
    \ve{G}_{\text{F}_1}(s)=\frac{1260}{s + 1260} \ve{I}_3
\end{equation}
are investigated. The filter ensures that the input-output map from $\ve{d}$ to $\ve{p}$ is proper. The filter bandwidth is chosen well above the closed loop bandwidths to minimize the influence on the pointing covariances. The calculation using the matrix exponential method presented in Section \ref{ss:MES} required a total of $60\,$ms on a standard laptop with $4.51\,$ GHz and $24\,$ GB memory.
Figure~\ref{fig:Exposure} shows the $x$-axis entries of the covariance matrices over the exposure time. The $z$-axis entries are similar up to the first decimal, while the off-diagonal terms have equivalent behavior but are orders of magnitude lower. 
The covariances satisfy the balance equation, Eq.~\eqref{eq:acov}, for all values of $T$. The displacement covariance is the main contributor to accuracy for exposure times $T\le2\,$ms. The smear contribution reaches a maximum at approximately $T = 3\,$ms. Jitter is the main contributor to pointing accuracy for $T\ge 100\,$ms.

\begin{figure}[ht!]
\centering
\begin{tikzpicture}
\definecolor{blue1}{RGB}{222,235,247}
\definecolor{blue2}{RGB}{158,202,225}
\definecolor{blue3}{RGB}{49,130,189}
%

\begin{groupplot}[group style={
                      	group name=myplot,
                      	group size= 2 by 1,
                        vertical sep=2.25cm,
                        horizontal sep = 1.75cm},
                      	height=0.3\columnwidth,
                      	width = 0.45\columnwidth,
                      	xmajorgrids=true,
			ymajorgrids=true,
			 grid style={densely dotted,white!60!black},
			  xmin = 0.1, xmax = 30,
			  ymin = -20, ymax = 5,
              xmode=log,
              legend pos= south east,
			   ]]

\nextgroupplot[	
				  ylabel= $\text{Sensitivity,}\,x\text{-axis}  \,{[\text{dB}]}$,
				 xlabel= $\text{Frequency}\, \omega \,{[\text{rad/s}]}$,
				 ]
\addplot[blue3, line width = 2.0] table[x expr = \thisrowno{0} ,y expr = \thisrowno{1} ,col sep=comma] {SensXX.csv};\label{pl:Nom1}
\addlegendentry{$\omega_\text{C}$}
\addplot[YellowOrange, line width = 2.0] table[x expr = \thisrowno{0} ,y expr = \thisrowno{2} ,col sep=comma] {SensXX.csv};\label{pl:Nom2}
\addlegendentry{$0.25\omega_\text{C}$}
\addplot [color=black, dashed, line width = 1pt]
	table[row sep=crcr]{%
		1.1 -100\\
		1.1  5 \\
	};
\addlegendentry{$\omega_\text{peak}$}

\nextgroupplot[	
				 ylabel= $\text{Sensitivity,}\,z\text{-axis}  \,{[\text{dB}]}$,
				 xlabel= $\text{Frequency}\, \omega  \,{[\text{rad/s}]}$,
				 ]
\addplot[blue3, line width = 2.0] table[x expr = \thisrowno{0} ,y expr = \thisrowno{1} ,col sep=comma] {SensYY.csv};\label{pl:Rob1}
\addplot[YellowOrange, line width = 2.0] table[x expr = \thisrowno{0} ,y expr = \thisrowno{2} ,col sep=comma] {SensYY.csv};\label{pl:Rob2}
\addplot [color=black, dashed, line width = 1pt]
	table[row sep=crcr]{%
		1.1 -100\\
		1.1  5 \\
	};
\addlegendentry{$\omega_\text{C}$}
\addlegendentry{$0.25\omega_\text{C}$}
\addlegendentry{$\omega_\text{peak}$}

\end{groupplot}
\end{tikzpicture}
\caption{Comparison of input sensitivity functions.}
\label{fig:BodeSens}
\end{figure}
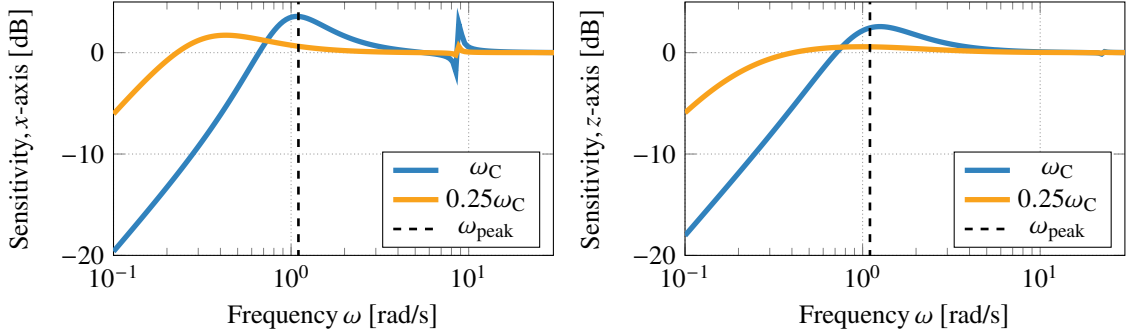

The second test case compares the pointing covariances of the nominal controller in Table~\ref{tab:Gains} and a controller achieving $25\%$ of the nominal tracking bandwidth $\omega_\text{C}$ using the filter 
\begin{equation}
 \ve{G}_{\text{F}_2}(s) = \frac{1260}{s + 1260} \frac{s^2+0.6736 s + 1.21}{s^2 + 0.1123 s + 1.21} \ve{I}_3.   
\end{equation}
The exposure time is $T=5\,$s.  
The disturbance filter $\ve{G}_{\text{F}_2}$ is composed of a low pass, equivalent to the first test case, and a peak filter with peak frequency $\omega_\text{peak} = 1.1\,$rad/s and nearly unit gain away from the peak frequency. The latter introduces a disturbance with a distinct frequency $\omega_\text{peak}$. Disturbances with such distinct frequency content are common in satellite pointing problems \cite{Ruth2010}.
Figure~\ref{fig:BodeSens} compares the input sensitivity function of the two closed loop systems about the $x$-axis and $y$-axis. Here, the sensitivity peaks of the nominal closed loop coincides with the resonance frequency $\omega_\text{peak}$. The sensitivity peaks of the slower controller are lower than the sensitivity peaks of the nominal controller at $\omega_\text{peak}$.
The covariance matrices for each controller are
\begin{align*}
\ve{\Sigma}_{\text{A},\omega_\text{C}} & = \bmtx  634.8 & -0.022 \\ -0.022 & 633.7 \emtx & \ve{\Sigma}_{\text{A},0.25\omega_\text{C}} & = \bmtx  632.4 & -0.004 \\ -0.004 & 632.3 \emtx\\
\ve{\Sigma}_{\text{D},\omega_\text{C}} & = \bmtx  0.210 &  0 \\ 0 & 0.177 \emtx & \ve{\Sigma}_{\text{D},0.25\omega_\text{C}} & = \bmtx  0.274 &  0 \\0 & 0.226 \emtx\\
\frac{1}{12}\ve{\Sigma}_{\text{S},\omega_\text{C}} & = \bmtx 2.004 & -0.006 \\ -0.006 &  1.572 \emtx & \frac{1}{12}\ve{\Sigma}_{\text{S},0.25\omega_\text{C}} & = \bmtx 1.149 & -0.002 \\ -0.002 &  1.108 \emtx \\
\ve{\Sigma}_{\text{J},\omega_\text{C}} & = \bmtx 632.6 & -0.017 \\ -0.017 & 632.0 \emtx & \ve{\Sigma}_{\text{J},0.25\omega_\text{C}} & = \bmtx 630.9 & -0.002 \\ -0.002 & 631.0 \emtx.
\end{align*}
The effect of the sensitivity peak manifests in the covariances. The slower controller, without a sensitivity peak overlap with $\omega_\text{peak}$, provides only slightly better accuracy and smaller jitter covariance, but the displacement covariance is slightly greater. The smear covariance is improved by the slower controller. However, the displacement and smear covariance are a small portion of the accuracy, which is dominated by jitter. Pushing the bandwidth higher in this example leads to reduced pointing performance. Sensor noise, which is not considered here, would have the opposite effect. Such behavior is observed in complex pointing scenarios \cite{Ruth2010}.
In general, there is a trade-off between low frequency disturbance rejection
and disturbance rejection around the closed-loop design bandwidth. Such test cases provide insight into pointing performance under broadband noise, which is beneficial for the iterative control design processes.

\section{Conclusion}
This Note presents a method to compute the displacement, smear, and jitter covariance matrices for a stable LTI pointing control system driven by white noise. The covariance matrices are computed from the solution of an augmented Lyapunov differential equation (LDE). Expressing the LDE in block state space form facilitates an efficient numerical calculation of these covariances using a matrix exponential. A previous method defines jitter as the sum of the smear and the jitter defined here, but it is important to distinguish smear and jitter, since they have different effects on image
quality and are specified separately in optical payload performance requirements. Our results,
however, do not treat the effect of other types of image motion such as quadratic, cubic, exponential,
and tonal, but these may be usually driven mostly by disturbances other than white noise.

\section*{Appendix}
The satellite dynamics are described by
\begin{equation}
\begin{split}
\left[\begin{array}{c} \ve{\dot{x}}_\text{r} \\ \hline \ve{\dot{x}}_\text{f}\end{array}\right]
    =& \left[\begin{array}{c|c} 
\ve{A}_\text{r} & \ve{A}_\text{rf} \\ \hline
\ve{0} & \ve{A}_\text{f} 
\end{array}\right]\left[\begin{array}{c} \ve{{x}}_\text{r} \\ \hline \ve{{x}}_\text{f}\end{array}\right] + \left[\begin{array}{c} \ve{B}_\text{r}\\ \hline \ve{B}_\text{f}\end{array}\right]\tau\\
\ve{y} = & \left[\begin{array}{c|c} \ve{C}_\text{r} & \ve{0} \end{array}\right]\left[\begin{array}{c} \ve{{x}}_\text{r} \\ \hline \ve{{x}}_\text{f}\end{array}\right],
\end{split}
\end{equation}
with
\begin{equation*}
\begin{split}
    \ve{A}_\text{r} &= \bmtx 0&0&0&0&0&	0\\[-10pt]
                       1&	0&	0&	0&	0&	0\\[-10pt]
                       0&	0&	0&	0&	0&	0\\[-10pt]
                       0&	0&	1&	0&	0&	0\\[-10pt]
                       0&	0&	0&	0&	0&	0\\[-10pt]
                       0&	0&	0&	0&	1&	0 \emtx,
\end{split}
\end{equation*}
\begin{equation*}
\ve{B}_\text{r} =\bmtx 0.004&-6.56\cdot 10^{-6}&-8.31\cdot 10^{-5}\\[-10pt]
0&	0&	0\\[-10pt]
-6.56\cdot 10^{-6}&0.001&1.54\cdot 10^{-5}\\[-10pt]
0&0&0\\[-10pt]
-8.31\cdot 10^{-5}&1.54\cdot 10^{-5}&0.012\\[-10pt]
0&0&0\emtx
\end{equation*}
    
\begin{equation*}
    \begin{split}
        \ve{C}_\text{r} &= \bmtx0&1&0&0&0&0\\[-10pt]
0&0&0&1&0&0\\[-10pt]
0&0&0&0&0&1\emtx.
    \end{split}
\end{equation*}

\begin{equation*}
\begin{split}
    \ve{A}_\text{f}=\left[\begin{array}{cccccc}
0 & 0 & 0 & 1 & 0 & 0 \\[-10pt]
0 & 0 & 0 & 0 & 1 & 0 \\[-10pt]
0 & 0 & 0 & 0 & 0 & 1 \\[-10pt]
-57.41 & 6.01 & -286.77 & -0.10 & 0.003 & -0.08 \\[-10pt]
0.51 & -436.86 & 9.84 & 9.10^{-4} & -0.23 & 0.003 \\[-10pt]
-7.18 & 2.93 & -1335.34 & -0.013 & 0.002 & -0.38 \\[-10pt]
0 & 0 & 0 & 0 & 0 & 0 \\[-10pt]
0 & 0 & 0 & 0 & 0 & 0 \\[-10pt]
0 & 0 & 0 & 0 & 0 & 0 \\[-10pt]
24.36 & 2.43 & 247.96 & 0.044 & 0.001 & 0.07 \\[-10pt]
-0.51 & 64.37 & -9.84 & -9.04 .10^{-4} & 0.03 & -0.003 \\[-10pt]
6.20 & 1.98 & 59.67 & 0.01 & 0.001 & 0.02
\end{array}\right. \ldots\\
\ldots\left.\begin{array}{cccccc}
0 & 0 & 0 & 0 & 0 & 0 \\[-10pt]
0 & 0 & 0 & 0 & 0 & 0 \\[-10pt]
0 & 0 & 0 & 0 & 0 & 0 \\[-10pt]
24.37 & -6.01 & 247.71 & 0.04 & -0.003 & 0.07 \\[-10pt]
0.20 & 64.37 & 6.65 & 3.7 \cdot 10^{-4} & 0.03 & 0.002 \\[-10pt]
6.21 & -2.93 & 59.67 & 0.01 & -0.002 & 0.027 \\[-10pt]
0 & 0 & 0 & 1 & 0 & 0 \\[-10pt]
0 & 0 & 0 & 0 & 1 & 0 \\[-10pt]
0 & 0 & 0 & 0 & 0 & 1 \\[-10pt]
-57.37 & -2.43 & -286.1 & -0.10 & -0.001 & -0.08 \\[-10pt]
-0.20 & -436.86 & -6.65 & -3.66 \cdot 10^{-4} & -0.23 & -0.002 \\[-10pt]
-7.16 & -1.98 & -1335.1 & -0.01 & -0.001 & -0.38
\end{array}\right],
\end{split}
\end{equation*}

\begin{equation*}
\begin{split}
   \ve{A}_\text{rf} = \left[\begin{array}{cccccc}
-1.67 & 0.12 & -17.75 & -0.003 & 6.16 \cdot 10^5 & -0.01 \\[-10pt]
0 & 0 & 0 & 0 & 0 & 0 \\[-10pt]
0.003 & -0.02 & 0.04 & 5.75 \cdot 10^{-6} & -1.14 \cdot 10^5 & 1.01 \cdot 10^{-5} \\[-10pt]
0 & 0 & 0 & 0 & 0 & 0 \\[-10pt]
0.13 & -16.76 & 2.56 & 2.35 \cdot 10^{-4} & -0.01 & 7.24 \cdot 10^{-4} \\[-10pt]
0 & 0 & 0 & 0 & 0 & 0
\end{array}\right.\ldots\\
\ldots \left.\begin{array}{cccccc}
1.67 & -0.12 & 17.72 & 0.003 & -6.16 \cdot 10^{-5} & 0.01 \\[-10pt]
0 & 0 & 0 & 0 & 0 & 0 \\[-10pt]
-0.003 & 0.02 & -0.03 & -5.32 \cdot 10^{-6} & 1.14 \cdot 10^5 & -8.48 \cdot 10^{-6} \\[-10pt]
0 & 0 & 0 & 0 & 0 & 0 \\[-10pt]
0.05 & 16.76 & 1.73 & 9.53 \cdot 10^{-5} & 0.01 & 4.89 \cdot 10^{-4} \\[-10pt]
0 & 0 & 0 & 0 & 0 & 0
\end{array}\right],
\end{split}
\end{equation*}

\vspace{-20pt}
\begin{equation*}
\begin{split}
\ve{B}_\text{f} &= \bmtx 0&	0&	0\\[-10pt]
0&0&0\\[-10pt]
0&0&0\\[-10pt]
0.05&-1.03\cdot 10^{-4}&-0.004\\[-10pt]
-3.19\cdot 10^{-4}&5.93\cdot 10^{-5}&0.05\\[-10pt]
0.01&-2.85\cdot 10^{-5}&-0.002\\[-10pt]
0&	0&	0\\[-10pt]
0&	0&	0\\[-10pt]
0&	0&	0\\[-10pt]
-0.05&	9.50\cdot 10^{-5}&-0.002\\[-10pt]
3.19\cdot 10^{-4}& -5.93\cdot 10^{-5}&-0.05\\[-10pt]
-0.01&2.40\cdot 10^{-5}&-1.38\cdot 10^{-4}\emtx
\end{split}
\end{equation*}

\vspace{-20pt}
\section*{Acknowledgments}

The first and second author acknowledge funding from NASA to develop 
the Image Motion Optical Transfer Function
and Pointing Performance Analysis
(IMOTF-PPA) Toolbox. The extended Bayard's method described in this Note is implemented as part of the Toolbox with minor differences. This Note was written independently of the Toolbox development. The last author acknowledges funding by the European Union under Grant No. 101153910. Views and opinions expressed are those
of the authors and do not necessarily reflect those of NASA or the European Union.

\vspace{-20pt}
\bibliography{ExtendedBayardsMethod}

\end{document}